\documentclass[lettersize,journal]{IEEEtran}
\usepackage{amsmath,amsfonts}
\usepackage{algorithmic}
\usepackage{algorithm}
\usepackage{array}
\usepackage[caption=false,font=normalsize,labelfont=sf,textfont=sf]{subfig}
\usepackage{textcomp}
\usepackage{stfloats}
\usepackage{url}
\usepackage{verbatim}
\usepackage{graphicx}
\usepackage{cite}
\begin{document}
	
\title{Hybrid Controller for Robot Manipulators in Task-Space with Visual-Inertial Feedback}
	
\author{Seyed Hamed Hashemi, Jouni Mattila
		% <-this % stops a space
\thanks{This work was supported by the Academy of Finland as part of the "High-precision autonomous mobile manipulators for future digitalized construction sites" project [Grant No. 335569].\\
The authors are with the Unit of Automation Technology and Mechanical Engineering, Faculty of Engineering and Natural Sciences, Tampere University, Tampere, Finland. (e-mail: hamed.hashemi@tuni.fi, jouni.mattila@tuni.fi)}% <-this % stops a space
\thanks{Manuscript received November 2023; revised }}
	
\markboth{,~Vol., No., November~2023}%
{Hashemi \MakeLowercase{\textit{et al.}}: Hybrid Controller for Robot Manipulators in Task-Space with Visual-Inertial Feedback}
	
	
\maketitle
	
\begin{abstract}
This paper presents a visual-inertial-based control strategy to address the task space control problem of robot manipulators. To this end, an observer-based hybrid controller is employed to control end-effector motion. In addition, a hybrid observer is introduced for a visual-inertial navigation system to close the control loop directly at the Cartesian space by estimating the end-effector pose. Accordingly, the robot's tip is equipped with an inertial measurement unit (IMU) and a stereo camera to provide task-space feedback information for the proposed observer. It is demonstrated through the Lyapunov stability theorem that the resulting closed-loop system under the proposed observer-based controller is globally asymptotically stable. Besides this notable merit (global asymptotic stability), the proposed control method eliminates the need to compute inverse kinematics and increases trajectory tracking accuracy in task-space. The effectiveness and accuracy of the proposed control scheme are evaluated through computer simulations, where the proposed control structure is applied to a 6 degrees-of-freedom long-reach hydraulic robot manipulator.
\end{abstract}
	
	
\section{Introduction}
The control problem of robot manipulators has been extensively studied over the past decades, and numerous control techniques have been developed to solve this issue \cite{spong2022historical}. This is because the demand for robotic manipulators has increased in industrial applications, such as welding, pick and place, manufacturing, and gripping. Hence, various controllers have been designed for robotic manipulators, which can be classified into two categories: task-space control schemes and joint-space control schemes. Most of the feedback techniques for the control of robot manipulators have been presented in joint-space \cite{liang2010adaptive}, while robot tasks are usually defined in Cartesian space. Both control techniques offer their own advantages and disadvantages, but in contrast to joint space control techniques, task-space control methods ensure asymptotically convergence of task-space tracking errors in the presence of model uncertainties, and they can compensate for elastic deformations within the gearbox \cite{molitor2023task} and eliminate the need to solve inverse kinematics.
	
Due to the aforementioned advantages, numerous studies have been dedicated to introducing controllers in Cartesian space. The first task-space controller was introduced in \cite{10.1115/1.3139651}, where the robot was directly controlled in the work-space and it was proven that the provided method is applicable when singularity or redundancy arises. Proportional-derivative (PD) control, along with a fully dynamic feed-forward control, were utilized in \cite{slotine1987adaptive} to control the manipulator in the absence of joint measurements. A cooperative control structure was introduced in \cite{zhao2022adaptive} to address the problem of task-space trajectory tracking of cooperative robotic manipulators. The proposed structure consists of a velocity observer, a cooperative control law, and a desired trajectory estimator, all of which were designed in Cartesian space. An adaptive Taylor series was employed in \cite{ahmadi2019task} to estimate the lumped uncertainty, such that the estimator parameters were adaptively tuned and the upper bound of the error was also obtained. To cope with uncertainties in kinematics and the gravity matrix, \cite{yazarel2002task} presented an adaptive feedback law in task-space for the set point control of a robotic manipulator. Based on the forward dynamics of robot manipulators, \cite{lee2020task} introduced a tracking control mechanism for manipulators in work-space, where the forward dynamics and impedance controller were used to overcome the inverse kinematics problem. The task-space control problem of robot manipulators in the presence of uncertainty in both dynamics and kinematics has been addressed in \cite{xiao2018exponential}, where two observers were presented to estimate uncertainties. In addition, it was shown that the proposed observer-based controller is globally and exponentially stable. The task-space control approaches provided in (\cite{10.1115/1.3139651,xiao2018exponential}) use only proprioceptive sensors, such as joint encoders and IMU, to provide feedback information. Consequently, the performance of these controllers depends highly on the inverse kinematic solution, which is subject to calibration errors and parameter uncertainty.
	
Due to recent advancements in sensor technology, a variety of control strategies in the task-space nowadays utilize extrospective sensors, such as stereo cameras, to provide task-space feedback information directly, enabling the elimination of an inverse kinematic solution. Vision-based controllers represent a category of task-space controllers that have gained significant attention in past decades \cite{su2023robotic}, because robots equipped with cameras can gather a greater amount of environmental data and are capable of handling the problems associated with unstructured environments. Vision-based controllers for robot manipulators are classified into two distinct classes: position-based visual servoing (PBVS) and image-based visual servoing (IBVS) \cite{cong2023review}. The former provides three-dimensional (3D) camera pose estimations of an object by utilizing visual features extracted from the image. In the latter case, the feedback control signals are designed on a 2D image plane. IBVS has some advantages over PBVS, such as robustness against calibration and modeling errors \cite{bechlioulis2019robust}. However, despite the advantages of PBVS and IBVS, they suffer from shortcomings, including sensitivity to calibration and local stability, as well as problems related to the interaction matrix (e.g., singularities and local minima) \cite{rastegarpanah2022optimized}. Furthermore, these methods are often formulated as kinematic control techniques, neglecting the impact of robot dynamics. 
	
To overcome the problems associated with IBVS and PBVS, \cite{li2013global} presented a novel vision-based controller, in whose structure the entire task-space has been divided into local regions, with a feedback law designed for each. The global stability of the proposed controller is guaranteed by the fusion of regional feedback laws, in consideration of robot dynamics. A visual-inertial-based controller was introduced in \cite{sandy2017dynamically} to control a hydraulically actuated arm in the presence of dynamic and unspecified base motion. In this control scheme, a Kalman-type filter utilized measurements provided by an IMU and a camera to estimate the end-effector motion of a controller. In order to provide Cartesian space feedback information, \cite{hashemi2023task} used the visual simultaneous localization and mapping (VSLAM) algorithm, which estimates the end-effector pose for a globally and asymptotically stable hybrid controller. 
	
Extending the aforementioned discussion, vision-based controllers that incorporate robot dynamics into their stability analyses offer greater advantages over IBVS and PBVS. Further, inspired by these control strategies, this paper realizes the advantages of the visual-inertial navigation algorithm provided in \cite{wang2020hybrid} and the hybrid controller described in \cite{hashemi2023task} and proposes a visual-inertial-based controller for robot manipulators. The paper's contributions can be outlined as follows.\\
$\bullet$ A hybrid observer is developed for visual-inertial navigation systems, which estimates the end-effector pose to provide task-space feedback information.\\
$\bullet$ A hybrid feedback law is employed to control the end-effector pose in the Cartesian space. In comparison to existing task-space control strategies that utilize only proprioceptive sensors, the proposed controller has a lower trajectory tracking error, as it uses an extrospective sensor.\\
$\bullet$ The main contribution is the global asymptotic stability of the resultant closed-loop system by considering the effects of robot dynamics.\\
	
This article is divided into six sections, beginning with the Introduction and ending with the Appendix. Section 2 provides preliminaries and notations, as well as a basic introduction to hybrid systems, visual-inertial kinematics and measurements, and dynamic and kinematic robot manipulator models. Thereafter, section 3 presents the proposed hybrid observer algorithm for visual-inertial navigation. The design and stability analysis of the proposed hybrid controller is described in section 4, and the proposed observer-based control strategy is validated utilizing simulation experiments in section 5. Finally, section 6 provides concluding remarks to summarize the paper.
	
\section{Preliminaries }
\subsection{Notations}
Let $\mathbb{R}$, $\mathbb{N}$, and ${{\mathbb{R}}^{n}}$ denote a set of real numbers, a set of natural numbers, and $n$-dimensional Euclidean space, respectively. $\mathbb{S}^n:=\{y\in \mathbb{R}^{n+1}: \|y\|=1 \} $ refers to a unit $n$-dimensional sphere and $\mathbb{B} :=\{y\in \mathbb{R}^n: \|y\| \le1 \}$ to a closed unit $n$-dimensional ball. $\left\| x \right\|=\sqrt{\left\langle x,x \right\rangle }$ represents the Euclidean norm of a vector $x \in {{\mathbb{R}}^{n}}$, where $\left\langle x,y \right\rangle :={{x}^{T}}y$ is the inner product and ${{\left\| x \right\|}_{\mathcal{A}}}:={{\min }_{y\in \mathcal{A}}}\left\| x-y \right\|$ is the shortest distance from $x$ to $\mathcal{A}$. Frobenius norm and skew-symmetric parts of a given matrix $A \in \mathbb{R}^{n \times n}$ are defined as $\left\| A \right\|_F=\sqrt{\left\langle A,A \right\rangle}=\sqrt{\text{tr}(A^{T}A)}$, and $\text{skew}(A)=(A-A^{T})/2$, respectively, where $\text{tr}(.)$ is the trace of $A$ and $\lambda(.)$ denotes its eigenvalue. This paper uses $SE(3):=\{ \mathbb{X}= \Psi(R,p):R \in SO(3), p \in \mathbb{R}^3\}$, and $SO(3):=\{R \in \mathbb{R}^{3\times3}:R^TR=RR^T=I, \text{det}(R)=1 \}$ to denote the special Euclidean group and the special orthogonal group of order three, respectively. For every $\mathbb{X}= \Psi(R,p)$, one has $\mathbb{X}^{-1}= \Psi(R^T,-R^Tp)$. Moreover, $\mathfrak{se}(3)$ represents the Lie algebra of $SE(3)$, which is given by:
	
\begin{equation*}\label{eq1}
	\begin{split}
		&\mathfrak{se}(3):=\{\mathbb{W}= \left[\begin{array}{c c} 
		\Gamma(\omega) & v \\
		0_{1\times3} & 0 
		\end{array}\right] :\omega, v \in \mathbb{R}^3 \}.\\
	\end{split}
\end{equation*}
The definition of frequently used maps is given as follows.
	
\begin{equation}\label{eq2}
	\begin{split}
	&\Gamma (y)=\left[ \begin{matrix}
		0 & -{{y}_{3}} & {{y}_{2}}  \\
		{{y}_{3}} & 0 & -{{y}_{1}}  \\
		-{{y}_{2}} & {{y}_{1}} & 0  \\
	\end{matrix} \right], \\
		& \varphi(A)=\frac{1}{2} \left[ \begin{matrix}
		A_{(3,2)}-A_{(2,3)} \\
		A_{(1,3)}-A_{(3,1)} \\
		A_{(2,1)}-A_{(1,2)} \\
		\end{matrix} \right], \ \bar{\varphi}(\mathbb{X})= \left[ \begin{matrix}
			\varphi(R) \\
			\frac{1}{2} p \\
		\end{matrix} \right], \\
		&\Psi(R,p)= \left[\begin{array}{c c} 
			R & p \\
			0_{1\times3} & 1\\
		\end{array}\right], \\
		&\Upsilon(B)=\Upsilon(\left[ \begin{matrix}
			A & B_2 \\
			B_3^T & B_4 \\
		\end{matrix} \right])= \left[ \begin{matrix}
			\text{skew}(A) & B_2 \\
			0_{n+1 \times 3} & 0_{n+1 \times n+1} \\
		\end{matrix} \right],\\
		& Ad_{\mathbb{X}} \mathbb{W} := \mathbb{X}\mathbb{W}\mathbb{X}^{-1}, \ \forall \mathbb{X} = \Psi(R,p)\\
		& y \in \mathbb{R}^{3 \times 1}, A \in \mathbb{R}^{3 \times3}, (B_2,B_3) \in \mathbb{R}^{3 \times n+1}, \\
		& B_4 \in \mathbb{R}^{n+1 \times n+1} \\
	\end{split}
\end{equation}
	
The tangent space of $SE(3)$ is represented by $T_{\mathbb{X}} SE(3):=\{\mathbb{X}\mathbb{W}|\mathbb{X} \in SE(3), \mathbb{W} \in \mathfrak{se}(3) \}$. Let $m:SE(3) \to \mathbb{R}$ be a differentiable smooth function, where the gradient of $m$ is denoted by $\nabla_{\mathbb{X}}m \in T_{\mathbb{X}} SE(3)$ and is computed as: 
	
\begin{equation}\label{eq3}
	dm.\mathbb{X}\mathbb{W}= \left\langle \nabla_{\mathbb{X}}m,\mathbb{X}\mathbb{W} \right\rangle_\mathbb{X}=\left\langle \mathbb{X}^{-1} \nabla_{\mathbb{X}}m,\mathbb{W} \right\rangle.
\end{equation}
	
In the above equation, $dm$ is the differential of $m$ and $\left\langle.,.\right\rangle_\mathbb{X}$ is the Riemannian metric on the matrix Lie group. The rotation matrix $R \in SO(3)$ can be parametrized with respect to the rotation angle $\theta \in \mathbb{R}$ and the rotation axis $y\in {\mathbb{S}^{2}}$ using the Rodrigues formula $\Re :\mathbb{R}\times {\mathbb{S}^{2}}\to SO(3)$, which is defined by: 
	
\begin{equation}\label{eq5}
	\begin{split}
		&\Re (\theta ,y)=I+\sin (\theta )\Gamma (y)+(1-\cos (\theta )){{\Gamma }^{2}}(y), \quad \text{or} \\
		&\Re (\theta ,y)=\exp(\theta \Gamma(y)).\\
	\end{split}
\end{equation}
	
\subsection{Hybrid System Framework}
This paper considers the hybrid dynamic system $\mathcal{H}$ of the form \cite{hashemi2021observer}:
	
\begin{equation}\label{eq6}
	\mathcal{H}:\left\{ \begin{matrix}
		\dot{x}=f(x,u), & (x,u)\in C \\
		{{x}^{+}}=g(x,u), & (x,u)\in D \\
	\end{matrix} \right.
\end{equation} 
where $x \in \mathbb{R}^n$ is the state vector and $u \in \mathbb{R}^m$ is the input signal. In the above framework for the hybrid dynamic system, $f:{{\mathbb{R}}^{n}}\times {{\mathbb{R}}^{m}}\to {{\mathbb{R}}^{n}}$, $g:{{\mathbb{R}}^{n}}\times {{\mathbb{R}}^{m}}\to {{\mathbb{R}}^{n}}$, $C\subset {{\mathbb{R}}^{n}}\times {{\mathbb{R}}^{m}}$, and $D\subset {{\mathbb{R}}^{n}}\times {{\mathbb{R}}^{m}}$ are referred to as the flow map, jump map, flow set, and jump set, respectively. In addition, $f$ defines the continuous dynamics of $C$, whereas $g$ describes the discrete dynamics of $D$. The jump set depicts where jumps may occur, and the flow set specifies where continuous evolution is permitted. A solution to the hybrid system is characterized in a hybrid time domain $E \subset {{\mathbb{R}}_{\ge 0}}\times \mathbb{N}$. A set $E$ is said to be a hybrid time domain if $E=\bigcup\limits_{i=1}^{I}{(\left[ {{t}_{i}},{{t}_{i+1}} \right],i)}$ for limited sequences of times $0={{t}_{0}}\le {{t}_{1}}\cdots \le {{t}_{I+1}}$. 
	
\textbf{Lemma 1} \cite{teel2012lyapunov}: A closed set $\mathcal{A} \subset \mathbb{R}^n$ is locally exponentially stable for $\mathcal{H}$ if there exists a continuously differentiable function $V:\mathbb{R}^n \to \mathbb{R}_{\ge0}$ such that for all $(\alpha_1 > \alpha_2, s_1, s_2, n)\in \mathbb{R}_{\ge0}$ satisfies:
	
\begin{equation}\label{eq7}
	\begin{split}
		& \alpha_2{\left\| x \right\|}_{\mathcal{A}}^n \le V(x) \le \alpha_1{\left\| x \right\|}_{\mathcal{A}}^n, \\
		& \forall x \in (C \cup D \cup g(D)) \cap (\mathcal{A}+s_1\mathbb{B})\\
		&\left\langle \nabla V(x),f \right\rangle \le -s_2V(x), \quad \forall x \in C \cap (\mathcal{A}+s_1\mathbb{B})\\
		& V(g) \le \exp(-s_2)V(x), \quad \forall x \in D \cap (\mathcal{A}+s_1\mathbb{B}).\\
	\end{split}
\end{equation}
	
Here, $V$ is defined in an open set that includes the closure of $C$. The set $\mathcal{A}$ is globally exponentially stable for $\mathcal{H}$ if $s_1 \to \infty$, and it is globally asymptotically stable for $\mathcal{H}$ if  $s_2 \to 0$ and $s_1 \to \infty$.
	
\textbf{Lemma 2} (Ruhe's trace inequality) \cite{ruhe1970perturbation}: For positive semidefinite Hermitian matrices $(A,B) \in \mathbb{R}^{n \times n}$, the following inequality holds:
	
\begin{equation*}
	\sum_{i=1}^{n}\lambda_i(A)\lambda_{n-i+1}(B) \le \text{tr}(AB) \le \sum_{i=1}^{n}\lambda_i(A)\lambda_i(B)
\end{equation*}
where $\lambda_1(.) \ge \ldots \ge \lambda_n(.) \ge 0$ (for both A,B).
	
\subsection{Kinematic Model and Measurements}
Kinematics of motion of a rigid body navigating 3D space is described, as follows:
	
\begin{equation}\label{eq8}
	\begin{split}
		& \dot{R}=R\Gamma (\omega ),\\
		& \dot{p}=v,\\
		& \dot{v}=Ra+g,\\
	\end{split}
\end{equation}
In equations (\ref{eq8}), $v \in \mathbb{R}^3$, and $p \in \mathbb{R}^3$ represent the linear velocity and position of the rigid body with respect to the inertial frame, respectively, and $g \in \mathbb{R}^3$ is the gravity vector. $\omega \in \mathbb{R}^3$ and $a \in \mathbb{R}^3$ denote the angular rate and acceleration obtained from the IMU in the body-fixed frame, respectively, and the kinematics (\ref{eq8}) can be compactly rephrased as follows:
	
\begin{equation}\label{eq11}
	\begin{split}
		& \dot{\mathcal{X}}=\mathcal{X}\mathcal{U}+G\mathcal{X}, \\
		& \mathcal{U}=\left[ \begin{matrix}
			\Gamma(\omega) & a & 0  \\
			0 & 0 & 1  \\
			0 & 0 & 0  \\
		\end{matrix} \right], \ G =\left[ \begin{matrix}
			0 & g & 0  \\
			0 & 0 & -1  \\
			0 & 0 & 0  \\
		\end{matrix} \right], \\ 
		& \mathcal{X}:=\bar{\Psi}(R,v,p)=\left[ \begin{matrix}
			R & v & p  \\
			0 & 1 & 0  \\
			0 & 0 & 1  \\
		\end{matrix} \right], \\
	\end{split}
\end{equation}
	
In the above equation, $\mathcal{X}$ belongs to the extended special Euclidean group of order 3, given by $SE_2(3):=\{\bar{\Psi}(R,v,p) | R\in SO(3), v,p \in \mathbb{R}^3\}$. Note that $\mathcal{X}^{-1}=\bar{\Psi}(R^T,-R^Tv,-R^Tp) \in SE_2(3)$. Besides measuring the acceleration and angular velocity, a stereo camera system mounted on the robot provides landmark measurements. Let $p_i \in \mathbb{R}^3$ denote the landmark positions in an inertial frame. It has been demonstrated that \cite{hamel2017riccati}, the 3D landmark position measurements expressed in the robot frame $y_i=R^T(p_i-p)$, can be determined by utilizing bearing measurements produced by a stereo camera. From (\ref{eq11}), one can deduce:
	
\begin{equation}\label{eq12}
	\beta_i:=\mathcal{X}^{-1}r_i=\left[\begin{array}{c} 
		y_i \\
		0 \\
		1 
	\end{array}\right], \ r_i=\left[\begin{array}{c} 
		p_i \\
		0 \\
		1 
	\end{array}\right]. 
\end{equation}
	
\textbf{Assumption} \cite{vasconcelos2010nonlinear}: It is assumed that at least three ($n \ge 3$) landmarks are available for measurements that are non-collinear.
	
\subsection{Dynamic and Kinematic Models of a Robot Manipulator}
The Euler–Lagrange equations of motion for $n$-link robotic manipulators are expressed as follows.
	
\begin{equation}\label{eq13}
	M(q) \ddot{q}+C(q,\dot{q})\dot{q}+G(q)+F(\dot{q}) = \tau,
\end{equation}
In equation (\ref{eq13}), $q = [q_1,q_2, \cdots, q_n]^T \in \mathbb{R}^n$ denotes the joint position vector and $(\dot{q}, \ddot{q}) \in \mathbb{R}^n$ its velocity and acceleration, respectively. $M(q) \in \mathbb{R}^{n \times n}$ represents the inertia matrix, which is a symmetric, positive-definite matrix. In addition, $C(q,\dot{q}) \in \mathbb{R}^{n \times n}$ and $G(q) \in \mathbb{R}^n$ represent the Coriolis and centrifugal matrix and the vector of gravitational forces, respectively. $F(\dot{q}) \in \mathbb{R}^n$ is the friction vector and $\tau \in \mathbb{R}^n$ is the control torques applied to the joints. The relation between joint space and Cartesian space can be described by so-called forward kinematics, given as:
	
\begin{equation}\label{eq14}
	x_{end}(t) = h(q(t)),
\end{equation}
where $x_{end}$ contains the end-effector position and attitude in the task space. The relation between Cartesian space velocity and joint space velocity are defined through the Jacobian matrix $J$, which is obtained by differentiating forward kinematics with respect to time.
	
\begin{equation}\label{eq15}
	\dot{x}_{end}(t) = \frac{\partial h(q)}{\partial q}\dot{q}(t) = J\dot{q}(t), \ \text{or}, \ \mathcal{Z} = \left[ \begin{matrix}
		\omega \\
		v  \\
	\end{matrix} \right] = J\dot{q}(t) \\
\end{equation}
Here, $\omega, v$ denote end-effector angular and linear velocities, respectively.
	
\section{Visual–Inertial Algorithm}
In this section, the proposed hybrid observer for a visual-inertial navigation system is introduced. The task of the proposed observer is to estimate the tip pose of the manipulator via IMU and stereo camera measurements. Due to the remarkable improvement of computer vision technology, estimation algorithms that utilize vision measurements gained significant interest over the last decade \cite{chae2020robust}. For instance, visual simultaneous localization and mapping (VSLAM) \cite{hashemi2022global}, visual–inertial odometry \cite{von2022dm}, and vision-aided inertial navigation systems \cite{wang2021nonlinear} are vision-based estimation techniques. Because the state space of attitude, i.e., $SO(3)$, is a non-contractible manifold, continuous observers among the matrix Lie groups, including $SO(3)$, $SE(3)$, and $SE_2(3)$, cannot ensure global convergence \cite{bhat2000topological}. Hence, a hybrid observer has been introduced in \cite{wang2020hybrid} for a visual-inertial navigation system, which guarantees global exponential convergence. Consequently, the proposed observer is built on the prior work of \cite{wang2020hybrid}, where a hybrid mechanism has been used to overcome topological obstructions. By enhancing the algorithm described in \cite{wang2020hybrid}, the dynamics of the proposed hybrid observer are designed as follows: 
	
\begin{equation}\label{eq16}
\begin{split}
	&\begin{cases}
	\dot{\hat{\mathcal{X}}}=\hat{\mathcal{X}}\mathcal{U}+G\hat{\mathcal{X}}-\Delta\hat{\mathcal{X}}, & \hat{\mathcal{X}} \in C \\
	\dot{q}=0, \\
	\end{cases}\\ 
	&\begin{cases}
	\hat{\mathcal{X}}^+=\mathcal{X}_q, & \hat{\mathcal{X}} \in D \\
	q^+=\underset{q \in \mathcal{Q}}{\arg\min} \ \mathcal{E}(\mathcal{X}_q),
	\end{cases}\\
	& C:=\mathcal{E}(\hat{\mathcal{X}})-\min_{\mathcal{X}_q\in \mathcal{Q}} \mathcal{E}(\mathcal{X}_q)\le \delta, \\
	& D:=\mathcal{E}(\hat{\mathcal{X}})-\min_{\mathcal{X}_q\in \mathcal{Q}} \mathcal{E}(\mathcal{X}_q)\ge \delta, \\
	&\mathcal{X}_q = \bar{\Psi}(\Re(q\theta,\ell),0,0) \bar{\Psi}(\hat{R},\hat{v},\hat{p}), \quad q \in \mathbb{N}\\
	&\Delta= -\Upsilon((r-\hat{\mathcal{X}}b)r^TK),\ \text{or}, \\
	& \Delta= \Upsilon(\tilde{\mathcal{X}}^{-1}(I-\tilde{\mathcal{X}})MK), \ \mathcal{E}(\mathcal{X})=\sum_{i=1}^{n} \|r_i-\mathcal{X}\beta_i\|^2\\
\end{split}
\end{equation}
with an arbitrary constant $\delta \in \mathbb{R}_{>0}$, $r:=[r_1 \ r_2 \ldots r_n] \in \mathbb{R}^{5 \times n}$, $b:=[\beta_1 \ \beta_2 \ldots \beta_n] \in \mathbb{R}^{5 \times n}$, $M=rr^T$, and the observer gain $K$ to be designed later. Furthermore, the switching variable $q$ belongs to a nonempty finite set $\mathcal{Q}:=\{\mathcal{X}_q \in SE_2(3): q \in \mathbb{N}, \ell \in \mathbb{S}^2, \theta \in \mathbb{R}_{>0} \}$, in which $(\theta, \ell)$ are chosen arbitrarily. It is worth noting that the system (\ref{eq11}) and observer dynamics (\ref{eq16}), defined on $SE_2(3)$, are non-invariant. Let $\tilde{\mathcal{X}}=\mathcal{X}\hat{\mathcal{X}}^{-1}$ with $\tilde{R}=R\hat{R}^T, \ \tilde{v}=v-\tilde{R}\hat{v}, \ \tilde{p}=p-\tilde{R}\hat{p}$ be the right-invariant estimation error and using $\dot{\hat{\mathcal{X}}}^{-1}=-\hat{\mathcal{X}}^{-1}\dot{\hat{\mathcal{X}}}\hat{\mathcal{X}}^{-1}$, the dynamics of $\tilde{\mathcal{X}}$ are calculated as follows:
	
\begin{equation}\label{eq17}
	\begin{split}
		& \dot{\tilde{\mathcal{X}}}=\dot{\mathcal{X}}\hat{\mathcal{X}}^{-1}+\mathcal{X}\dot{\hat{\mathcal{X}}}^{-1} \\
		& \Rightarrow \dot{\tilde{\mathcal{X}}}=G\tilde{\mathcal{X}}-\tilde{\mathcal{X}}G+\tilde{\mathcal{X}}\Delta \\
	\end{split}
\end{equation}
\textbf{Theorem 1}: Consider the visual-inertial kinematics (\ref{eq11}) evolving on the matrix Lie group $SE_2(3)$. The observer provided in (\ref{eq16}) guarantees the global asymptotic convergence of $\tilde{\mathcal{X}}$ to $I_{5 \times 5}$.\\
\textbf{Proof}: The candidate Lyapunov function is given by:
	
\begin{equation}\label{eq18}
	V_o = \frac{1}{2} \text{tr}((I-\tilde{\mathcal{X}})(I-\tilde{\mathcal{X}})^T).
\end{equation}
Taking the derivative of $V_o$ with respect to time and substituting $(\Delta, \dot{\tilde{\mathcal{X}}})$ from (\ref{eq16}) and (\ref{eq17}), one obtains:
	
\begin{equation}\label{eq19}
\begin{split}
	& \dot{V_o} = -\text{tr}(\dot{\tilde{\mathcal{X}}}(I-\tilde{\mathcal{X}})^T) \Rightarrow \\
	& \dot{V_o} = \text{tr}(\left[ \begin{matrix}
		(I-\tilde{R})g\tilde{v}^T+\tilde{v}\tilde{p}^T & (I-\tilde{R})g & \tilde{v}  \\
		0 & 0 & 0  \\
		0 & 0 & 0  \\
	\end{matrix} \right])\\
	&-\text{tr}((I-\tilde{\mathcal{X}})MK(I-\tilde{\mathcal{X}})^T) \Rightarrow \\
	& \dot{V_o} =  -(\text{tr}((I-\tilde{\mathcal{X}})MK(I-\tilde{\mathcal{X}})^T) \\ & -\text{tr}((I-\tilde{R})g\tilde{v}^T+\tilde{v}\tilde{p}^T)).\\
\end{split}
\end{equation}
The observer gain is designed as:
	
\begin{equation}\label{eq20}
K = M \mathcal{K} = M \left[ \begin{matrix}
	\bar{g}^2I_{3 \times 3} & 0 & 0  \\
	0 & k_{v1} & 0  \\
	0 & k_{v2} & k_p  \\
\end{matrix} \right],
\end{equation}
with $k_{v1},k_{v2},k_p \in \mathbb{R}_{> 0}$, which are arbitrary constants, and $\bar{g}=\|g\|$. It follows from Lemma 2 that:
	
\begin{equation}\label{eq21}
\begin{split}
	&\text{tr}((I-\tilde{\mathcal{X}})M^2\mathcal{K}(I-\tilde{\mathcal{X}})^T) \ge \\ &\text{tr}((I-\tilde{\mathcal{X}})\mathcal{K}(I-\tilde{\mathcal{X}})^T) \lambda_n (M^2). \\
\end{split}
\end{equation}
Consequently, to prove $\dot{V_o} \le 0$, it is sufficient to show that:
	
\begin{equation}\label{eq22}
\begin{split}
	&\text{tr}((I-\tilde{\mathcal{X}})\mathcal{K}(I-\tilde{\mathcal{X}})^T) \lambda_n (M^2) \ge \\
	& \text{tr}((I-\tilde{R})g\tilde{v}^T+\tilde{v}\tilde{p}^T). \\
\end{split}
\end{equation}
The above equation can be further simplified as follows.
	
\begin{equation}\label{eq23}
\begin{split}
	& \lambda_n (M^2)(\bar{g}^2\left\| (I-\tilde{R}) \right\|^2_F+k_{v1}\left\| \tilde{v} \right\|^2+k_{v2}\langle\,\tilde{v},\tilde{p}\rangle \\
	&+k_p\left\| \tilde{p} \right\|^2) \ge \langle\,(I-\tilde{R})g,\tilde{v}\rangle+\langle\,\tilde{v},\tilde{p}\rangle \\
\end{split}
\end{equation}
By utilizing the fact that $\left\| Ax \right\| \le \left\| A \right\|_F \left\| x \right\|$ for all $A \in \mathbb{R}^{n \times n}$ and $x \in \mathbb{R}^n$, one has:
	
\begin{equation}\label{eq24}
\begin{split}
	& \lambda_n (M^2)(\left\| (I-\tilde{R})g \right\|^2_F+k_{v1}\left\| \tilde{v} \right\|^2+k_{v2}\langle\,\tilde{v},\tilde{p}\rangle \\
	&+k_p\left\| \tilde{p} \right\|^2) \ge \langle\,(I-\tilde{R})g,\tilde{v}\rangle+\langle\,\tilde{v},\tilde{p}\rangle. \\
\end{split}
\end{equation}
It can easily be shown that:
	
\begin{equation}\label{eq25}
	\left\| (I-\tilde{R})g-\tilde{v} \right\|^2+\left\| \tilde{p}+\tilde{v} \right\|^2 \ge 0,
\end{equation}
which completes the proof. $\square$
	
\section{Proposed Control Law}
The following section presents the proposed hybrid controller. Due to the existence of a set of Lebesgue measuring zero on the tangent bundle of the systems evolving on $SO(3)$, conventional continuous and discontinuous controllers cannot globally stabilize such a system to the desired reference set \cite{hashemi2021global}. Therefore, hybrid feedback laws have been designed to control systems evolving on matrix Lie groups, including $SO(3)$ \cite{hashemi2023observer}, $\mathbb{S}^3$ \cite{hashemi2021quaternion}, and $SE(3)$ \cite{wang2021hybrid}. To the best of the authors' knowledge, the first hybrid controller for robot manipulators in task-space has been introduced in \cite{hashemi2023task}. This controller was directly designed in end-effector space, denoted as $SE(3)$, by taking the gradient of a potential function. Nonetheless, the main drawback of this control technique is that it might create a severe rotation. Hence, this paper introduces a hybrid feedback law that adopts the advantages of the controller described in \cite{hashemi2023task} while also overcoming its shortcoming.
	
It has been proven that $1/2 \text{tr}(I-R^T_dR)$ is locally positive-definite when the angle between the actual attitude $R$ and the desired attitude $R_d$ is less than $180^{\circ}$ \cite{bullo2005geometric}. Based on this fact, the potential function $\mathbb{U}: SE(3) \times \mathbb{R} \rightarrow \mathbb{R}$ is given by
	
\begin{equation}\label{eq26}
\begin{split}
	&\mathbb{U}(\mathbb{X},h)=\frac{1}{2}\text{tr}((I-\mathbb{X})\mathbb{K}_c(I-\mathbb{X})^T), \\	
	&\mathbb{K}_c=k_c(I-\mathbb{X}_h), \ \mathbb{X}_h = \Psi(\Re (\theta_h ,\jmath_h),0),\\
\end{split}
\end{equation}
where $k_c=\text{diag}([k_{ci} \in \mathbb{R}_{\ge 0}])$ for $i=1 \ldots4$, and $(\jmath_h,\theta_h)$ belongs to a compact set $\Xi := \{\theta_h \in \mathbb{R}, \jmath_h \in \mathbb{S}^2: |\theta_h| \le \pi/10, \jmath_h \in \{[1 \ 0 \ 0]^T,[0 \ 1 \ 0]^T,[0 \ 0 \ 1]^T \}\}$. The gradient of $\mathbb{U}(\mathbb{X},h)$ is computed as follows by applying the gradient calculation method expressed in \cite{wang2021hybrid}:
	
\begin{equation}\label{eq27}
\bar{\varphi}(\mathbb{X}^{-1}\nabla_{\mathbb{X}}\mathbb{U}(\mathbb{X},h))=\bar{\varphi}((I-\mathbb{X}^{-1})\mathbb{K}_c).
\end{equation}
As a result, the proposed hybrid controller for robot manipulators for tracking task-space paths is designed as follows:
	
\begin{equation}\label{eq28}
\begin{split}
& \tau^* = N(q,\dot{q})-M(q)J^{-1}(\dot{J}\dot{q}+\bar{\varphi}(\mathbb{X}_e^{-1}\nabla_{\mathbb{X}_e}\mathbb{U}(\mathbb{X}_e,h)+K_d\mathbb{Y})), \\
& N(q,\dot{q}) = C(q,\dot{q})\dot{q}+G(q)+F(\dot{q}). \\
\end{split}
\end{equation}
Here, $K_{d} \in \mathbb{R}_{\ge 0}$ denotes the controller gain, and $(\mathbb{X}_e, \mathbb{Y})$ will be given later. In the above equations, $h$ is the switching variable, determined by the following hybrid mechanism.
	
\begin{equation}\label{eq29}
\begin{split}
	&\begin{cases}
	\dot{h}=0, & (\mathbb{X}_e,h) \in C' \\
	\end{cases}\\ 
	&\begin{cases}
	h^+=\underset{h' \in \Xi}{\arg\min} \ \mathbb{U}(\mathbb{X}_e,h'), & (\mathbb{X}_e,h) \in D' \\
	\end{cases}\\
	& C':=\{(\mathbb{U}(\mathbb{X}_e,h)-\min_{h' \in \Xi}  \mathbb{U}(\mathbb{X}_e,h')\le \delta), \\
	& D':=\{(\mathbb{U}(\mathbb{X}_e,h)-\min_{h' \in \Xi} \mathbb{U}(\mathbb{X}_e,h')\ge \delta), \\
\end{split}
\end{equation}
	
\textbf{Theorem 2}: The proposed hybrid feedback law (\ref{eq28}) makes the compact set $\mathcal{A}:=\{\mathbb{X}_d \in SE(3),\tilde{\mathcal{X}} \in SE_2(3) : \mathbb{X}_d = \bar{\Psi}(R_d,p_d), \tilde{\mathcal{X}}=I\}$ globally asymptotically stable for the closed-loop system with the proposed observer (\ref{eq16}).\\
\textbf{Proof}: Theorem 2 is proven in two steps based on Lemma 1.\\
\textbf{Step 1}: By considering the manipulator end-effector as a rigid body, its kinematic equation of motion can be expressed as:
	
\begin{equation}\label{eq30}
	\dot{\mathbb{X}} = \mathbb{X}\left[ \begin{matrix}
		\Gamma(\omega) & v \\
		0 & 0 \\
	\end{matrix} \right] =  \mathbb{X}\mathbb{W}.
\end{equation}
The trajectory-tracking error is described by $\mathbb{X}_e=\mathbb{X}_d^{-1}\mathbb{X}$. One can obtain the dynamics of $\mathbb{X}_e$ as:
	
\begin{equation}\label{eq31}
	\begin{split}
		& \dot{\mathbb{X}}_e = \dot{\mathbb{X}}_d^{-1}\mathbb{X}+\mathbb{X}_d^{-1}\dot{\mathbb{X}} \Rightarrow \\
		& \dot{\mathbb{X}}_e = \mathbb{X}_e(\mathbb{W}-Ad_{\mathbb{X}_e^{-1}}\mathbb{W}_d)=\mathbb{X}_e\mathbb{Y} \\
	\end{split}
\end{equation}
where $\dot{\mathbb{X}}_d^{-1}$ is replaced with $\dot{\mathbb{X}}_d^{-1}=-\mathbb{X}_d^{-1} \dot{\mathbb{X}}_d \mathbb{X}_d^{-1}$, and $\mathbb{W}_d$ belongs to $\mathfrak{se}(3)$ with the desired constant linear $v_d$ and angular $\omega_d$ velocities. Finally, the candidate Lyapunov function is defined by:
	
\begin{equation}\label{eq32}
	V(\mathbb{X}_e,\tilde{\mathcal{X}},\mathbb{Y}) = \mathbb{U}(\mathbb{X}_e,h)+V_o+\frac{1}{2}\bar{\varphi}^T(\mathbb{Y})\bar{\varphi}(\mathbb{Y}).
\end{equation}
One can easily find the derivative of $V$ with respect to time along the trajectories of (\ref{eq31}), as: 
	
\begin{equation}\label{eq33}
	\dot{V} = \left\langle \nabla_{\mathbb{X}_e}\mathbb{U},\mathbb{X}_e\mathbb{Y} \right\rangle_{\mathbb{X}_e}+\dot{V_o}+\bar{\varphi}^T(\mathbb{Y})\bar{\varphi}(\dot{\mathbb{Y}}).
\end{equation}
From equation (\ref{eq3}) and the fact that $\dot{\mathbb{W}}_d=0$, one has:

\begin{equation}\label{eq34}
	\dot{V} = \dot{V_o} + \left\langle \mathbb{X}_e^{-1}\nabla_{\mathbb{X}_e}\mathbb{U},\mathbb{Y} \right\rangle +\bar{\varphi}^T(\mathbb{Y})\bar{\varphi}(\dot{\mathbb{W}}).
\end{equation}
In view of (\ref{eq15}), one finds that $\bar{\varphi}(\mathbb{W}) = \mathcal{Z}$ and 
	
\begin{equation}\label{eq35}
	\dot{\mathcal{Z}} = J\ddot{q}+\dot{J}\dot{q}.
\end{equation}
by substituting (\ref{eq13}) into (\ref{eq35}), one has:
	
\begin{equation}\label{eq36}
	\dot{\mathcal{Z}} = JM^{-1}(q)(\tau-N(q,\dot{q}))+\dot{J}\dot{q}.
\end{equation}
One can obtain the following equation as a result of applying the proposed hybrid controller (\ref{eq28}).
	
\begin{equation}\label{eq37}
	\dot{\mathcal{Z}} = -\bar{\varphi}(\mathbb{X}_e^{-1}\nabla_{\mathbb{X}_e}\mathbb{U}+K_d\mathbb{Y})
\end{equation}
Accordingly, replacing (\ref{eq37}) with (\ref{eq34}), one shows that:
	
\begin{equation}\label{eq38}
	\begin{split}
		& \dot{V} = \dot{V_o}+\left\langle \mathbb{X}_e^{-1}\nabla_{\mathbb{X}_e}\mathbb{U},\mathbb{Y} \right\rangle \\
		& -K_d\bar{\varphi}^T(\mathbb{Y})\bar{\varphi}(\mathbb{Y}) - \bar{\varphi}^T(\mathbb{Y})\bar{\varphi}(\mathbb{X}_e^{-1}\nabla_{\mathbb{X}_e}\mathbb{U}).\\
	\end{split}
\end{equation}
It has been proven that $\dot{V_o} \le 0$; hence, one can deduce that $\dot{V} \le 0$. This implies that $\mathbb{X}_e$ and $\tilde{\mathcal{X}}$ are globally bounded. Likewise, $\ddot{V}$ is also globally bounded. Applying Barbalat's lemma, one concludes that $\lim_{t \to +\infty} \dot{V}=0$; therefore, $\lim_{t \to +\infty} \mathbb{X} \rightarrow \mathbb{X}_d$ and $\lim_{t \to +\infty} \mathcal{X} \rightarrow \hat{\mathcal{X}}$. \\
	
\begin{figure}
	\centering
	\includegraphics[width=1\linewidth]{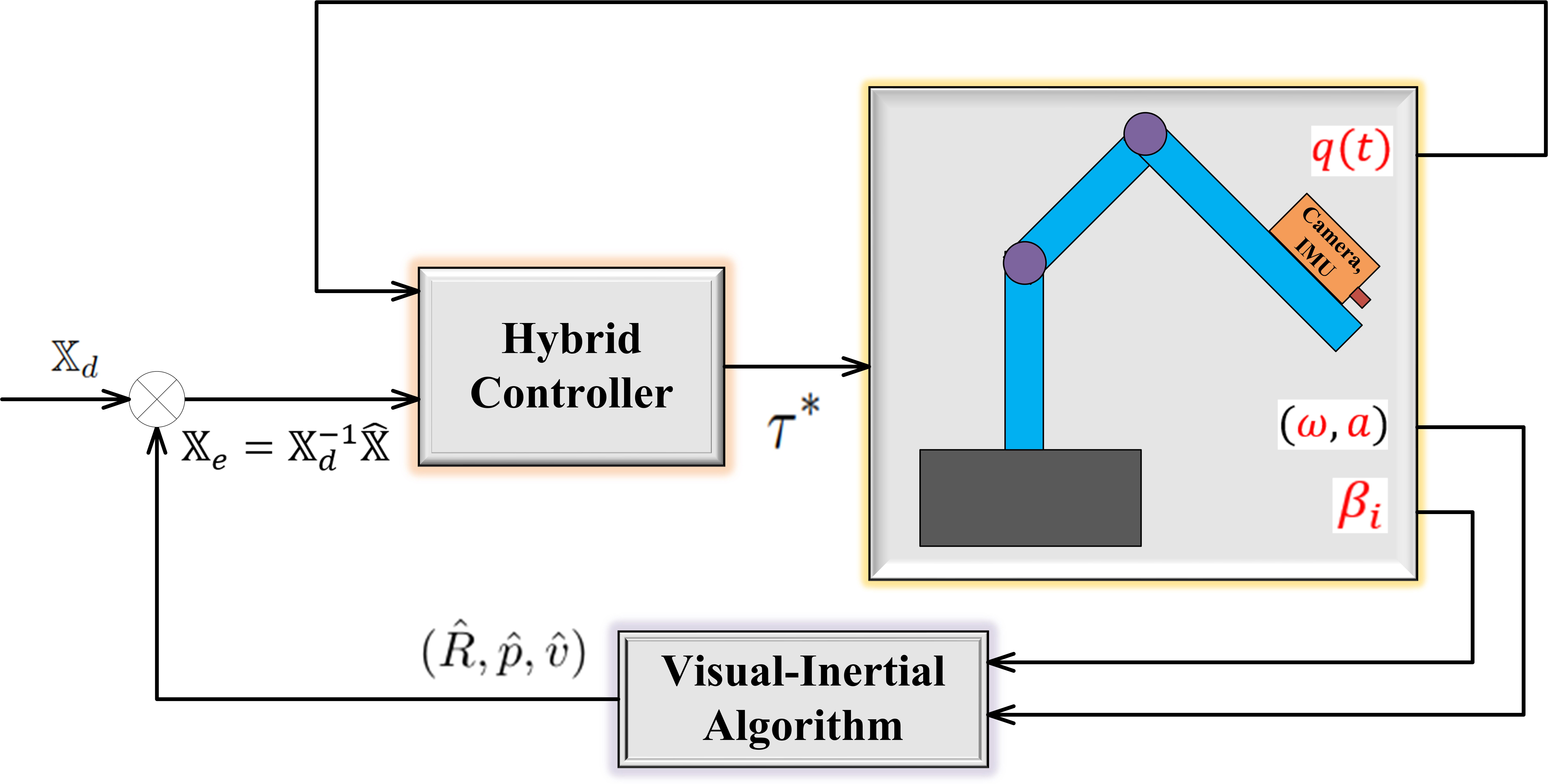}
	\caption{Sketch of the proposed visual-inertial-based controller.}
	\label{fig1}
\end{figure}

\textbf{Step 2}: In this step, it is demonstrated that the Lyapunov function stays in a negative state when switching variables $(h,q)$ shift towards the next values. Following equation (\ref{eq7}), the variation in $V(\mathbb{X}_e,\tilde{\mathcal{X}},\mathbb{Y})$ during jumps is given by:
	
\begin{equation}\label{eq39}
	\begin{split}
		& V^+(\mathbb{X}_e,\tilde{\mathcal{X}},\mathbb{Y})-V(\mathbb{X}_e,\tilde{\mathcal{X}},\mathbb{Y})= \\
		&\mathbb{U}(\mathbb{X}_e,h^+)+V_o(\mathcal{\tilde{X^+}})-\mathbb{U}(\mathbb{X}_e,h)-V_o(\mathcal{\tilde{X}}) \\
	\end{split}
\end{equation}
In view of (\ref{eq29}), (\ref{eq16}), and the fact that $\mathcal{E}(\hat{\mathcal{X}}) \ge V_o(\tilde{\mathcal{X}})$, one shows that:
	
\begin{equation}\label{eq40}
	\begin{split}
		& \min_{\mathcal{X}_q\in \mathcal{Q}} V_o(\mathcal{X}_q)-V_o(\mathcal{X}_q) \le -\delta, \\
		& \min_{h' \in \Xi} \mathbb{U}(\mathbb{X}_e,h') - \mathbb{U}(\mathbb{X}_e,h) \le -\delta. \\
	\end{split}
\end{equation}
This reveals that the Lyapunov function remains negative during jumps and completes the proof.  $\square$\\
It is important to mention that utilizing the estimated pose $\bar{\Psi}(\hat{R},\hat{p})$ to close the control loop still provides proof of stability, as $\lim_{t \to +\infty} \mathcal{X} \rightarrow \hat{\mathcal{X}}$. Figure (\ref{fig1}) shows the block diagram of the visual-inertial-based control scheme, and the salient features of this structure are 1) global asymptotic stability and 2) low computational time.
	
\section{Simulation Results}
\begin{figure}
	\centering
	\includegraphics[width=1\linewidth]{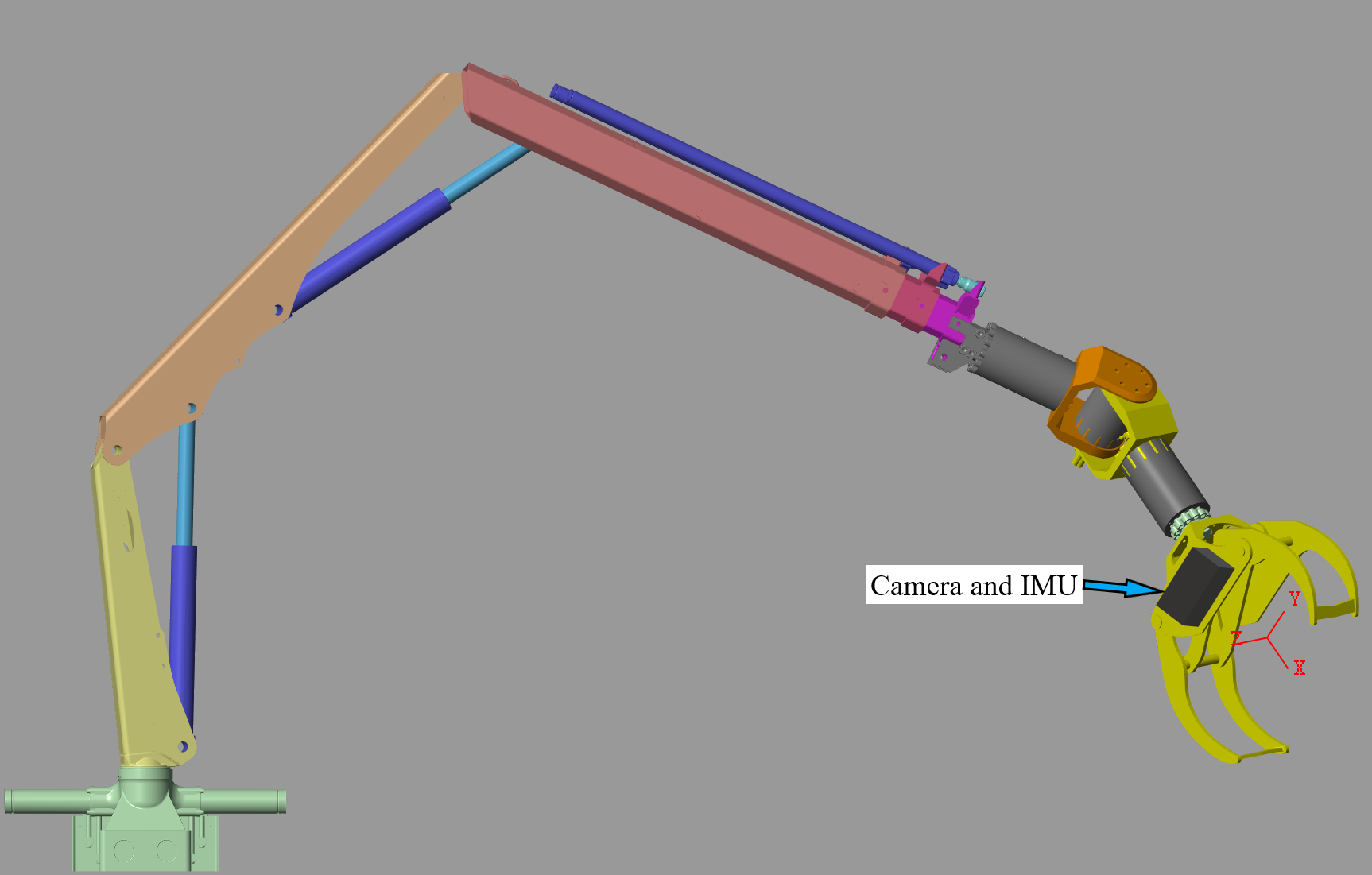}
	\caption{3-D hydraulic manipulator model in Simscape Multibody\textsuperscript{TM}.}
	\label{fig2}
\end{figure}

Based on numerical simulation results, this section peruses the performance of the visual-inertial-based controller. Simscape Multibody\textsuperscript{TM} is used to simulate a 6-degrees-of-freedom (DOF) hydraulic manipulator with a long arm and a gripper, demonstrated in Figure (\ref{fig2}). Due to their long reachability and high lifting capacity, hydraulic manipulators have attracted considerable attention from both academia and industry \cite{10177681}. Hence, in this paper, a full-scale commercial hydraulic manipulator is considered as a case study, whose further supplementary material can be found in \cite{petrovic2022mathematical}.
	
Robotic manipulation tasks are often given in Cartesian space, such as grasping, painting, and welding. Likewise, in this paper, the manipulator is controlled to draw a square with sides of 50 cm. Nonetheless, the majority of controllers for robotic manipulators are designed in joint space. In these controllers, robot tasks are converted from task-space to joint space through inverse kinematics. When faced with uncertainties in kinematics and dynamics, these controllers cannot guarantee precise path following in task-space. Hence, this paper suggests utilizing cameras to provide task-space feedback information $\beta_i$ that yields enhancements in trajectory-tracking precision. The joint measurements and end-effector pose estimated by the visual-inertial algorithm (\ref{eq16}) are then used as feedback data for the introduced control law (\ref{eq28}).
	
The observer is initialized at $[3.6 \ 1 \ 0]$, and it differs from the actual path to depict the robustness of the proposed visual-inertial algorithm against initial conditions. The controller and observer gains are calculated as follows, through trial and error until a satisfactory level of performance is achieved:
	
\begin{equation*}
\mathcal{K} = \left[ \begin{matrix}
	\bar{g}^2I_{3 \times 3} & 0 & 0  \\
	0 & 9 & 0  \\
	0 & 25 & 6  \\
\end{matrix} \right], \quad k_c = 300 I_{4 \times 4}, \quad K_d = 5.
\end{equation*}
	
\begin{figure}
	\centering
	\includegraphics[width=1\linewidth]{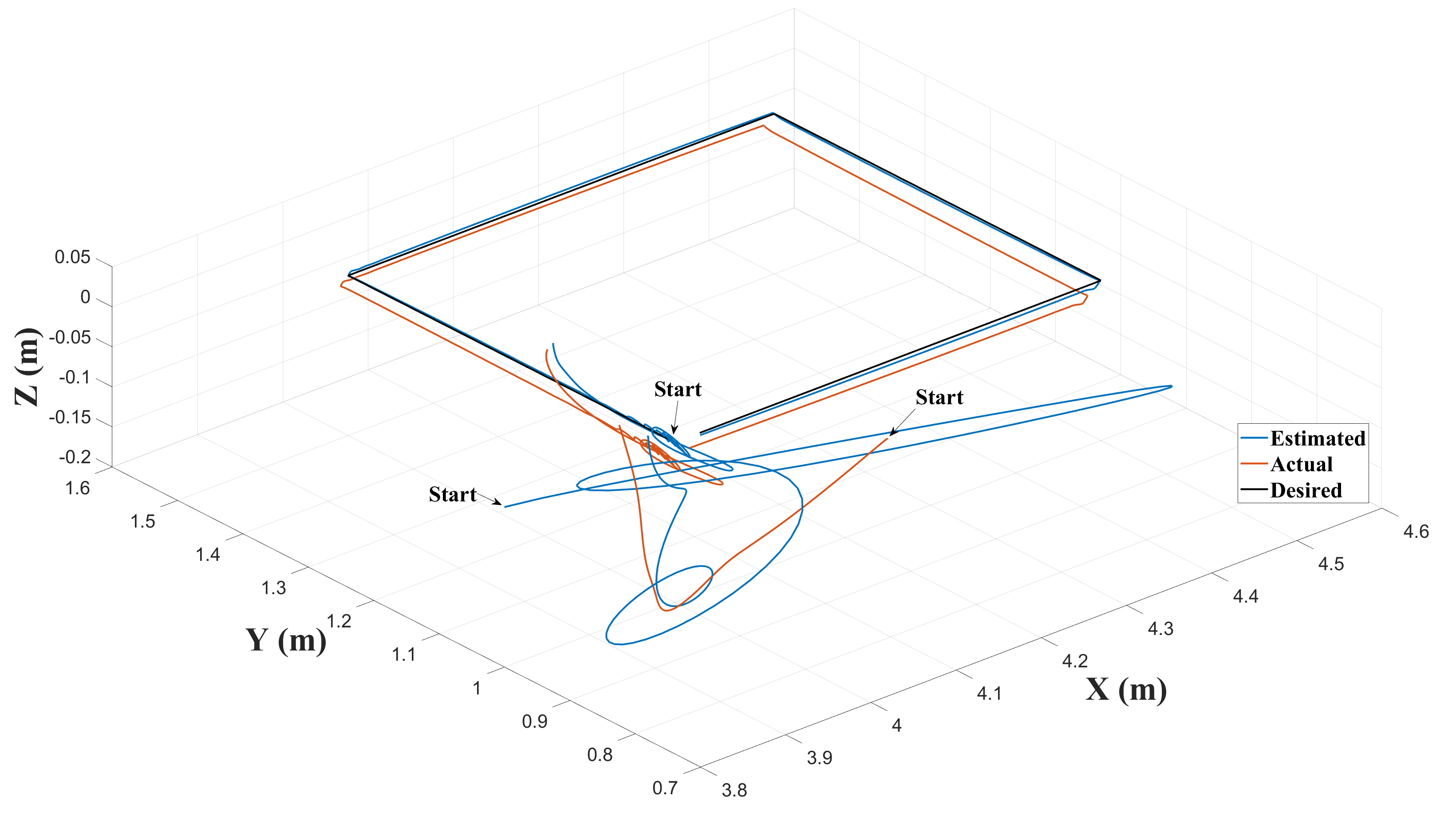}
	\caption{3D view of the desired, estimated, and actual trajectories.}
	\label{fig3}
\end{figure}

\begin{figure}
	\centering
	\includegraphics[width=1\linewidth]{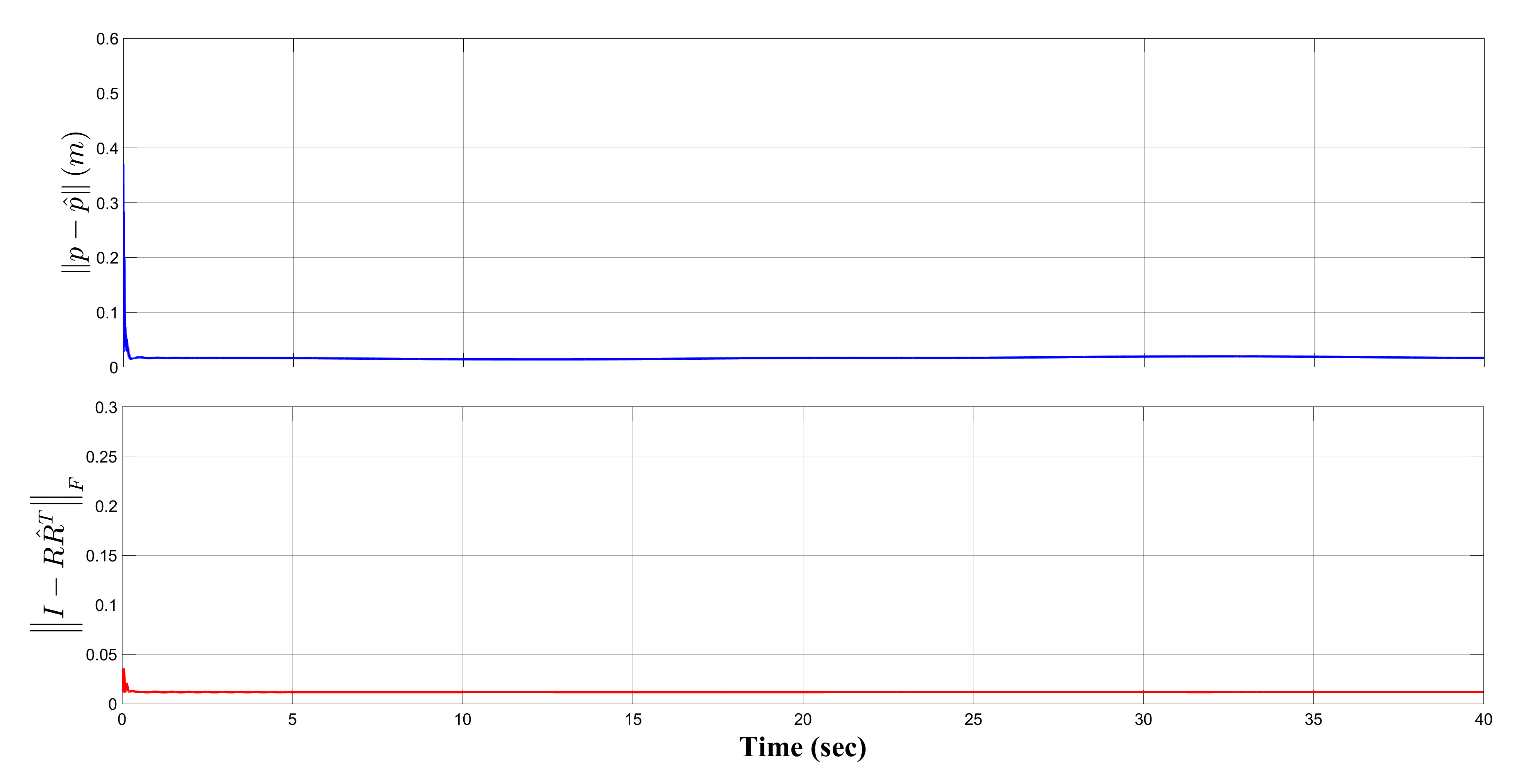}
	\caption{Time evolution of position and attitude estimation errors.}
	\label{fig4}
\end{figure}

\begin{figure}
	\centering
	\includegraphics[width=1\linewidth]{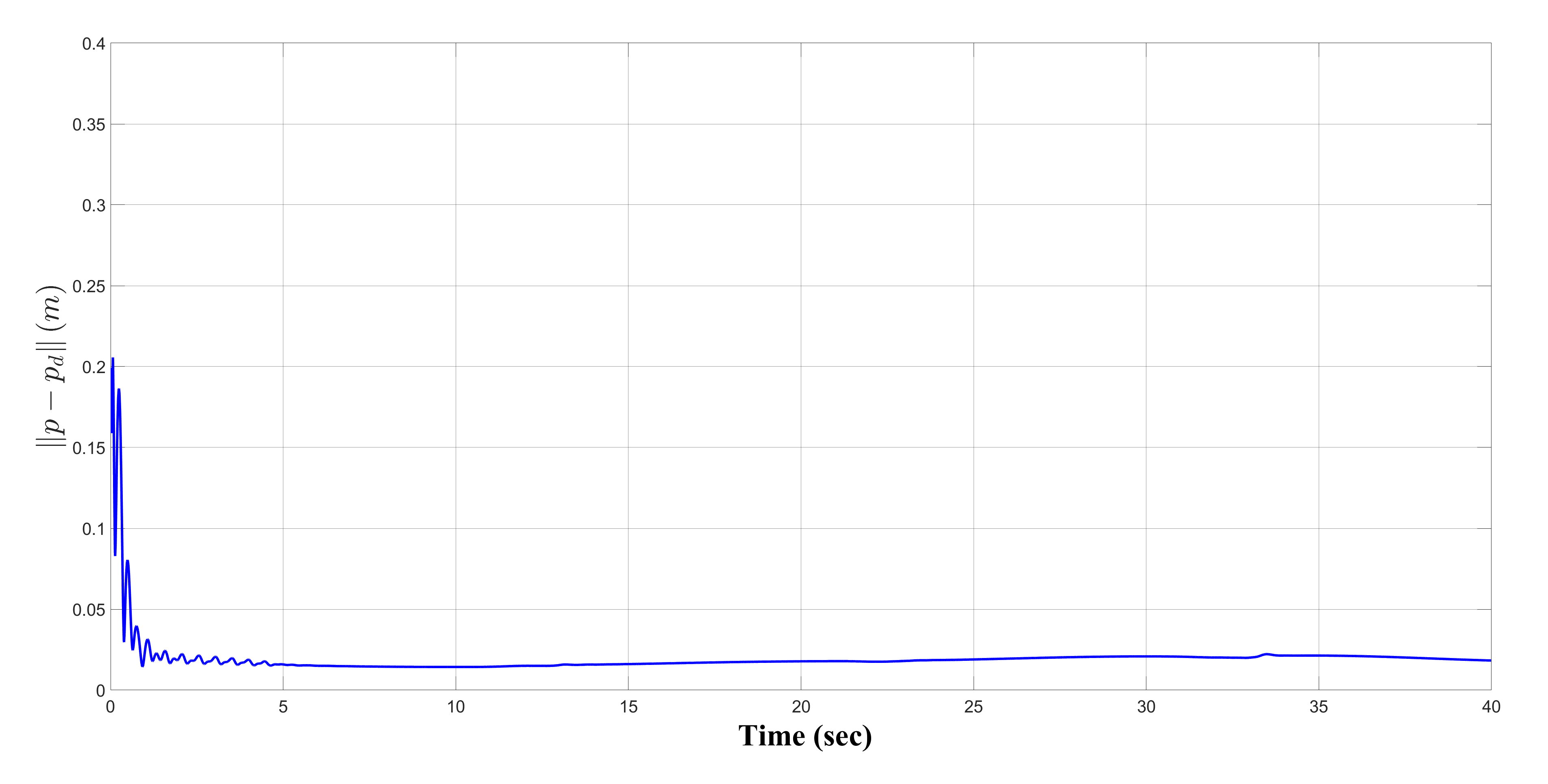}
	\caption{Task-space position tracking error.}
	\label{fig5}
\end{figure}

\begin{figure}
	\centering
	\includegraphics[width=1\linewidth]{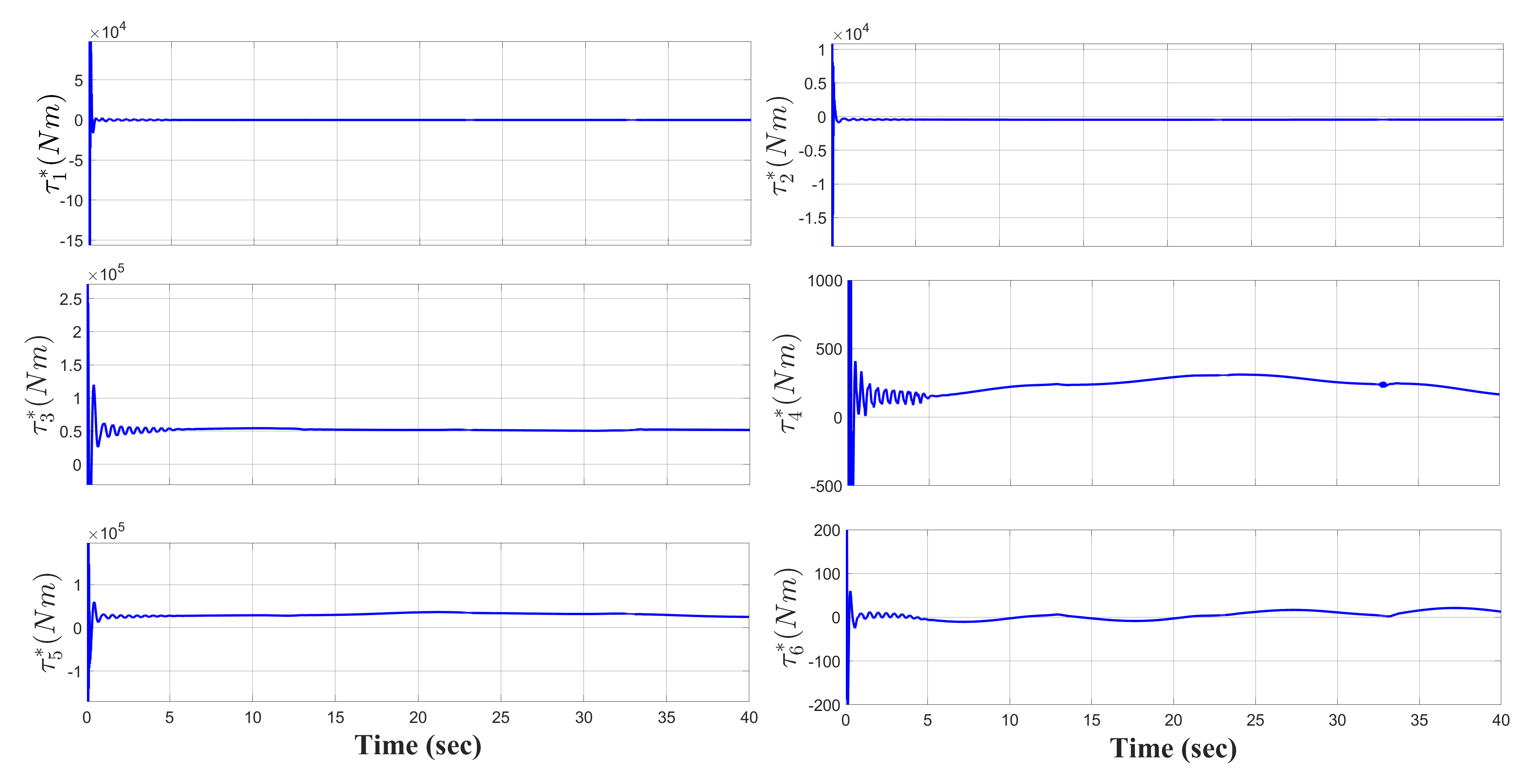}
	\caption{Time response of control torque.}
	\label{fig6}
\end{figure}

The simulation results presented in Figures (\ref{fig3}-\ref{fig6}) are obtained by applying the visual-inertial-based control strategy to the 6-DOF long-reach hydraulic manipulator. Figure (\ref{fig3}) shows a 3D view of the estimated path and trajectory drawn by the end-effector pose under the proposed observer-based controller tracking the desired path. This figure implies that the proposed visual-inertial-based control approach can track the desired trajectory within an acceptable error range. The evolution of estimation and tracking errors versus time is depicted in Figure (\ref{fig4},\ref{fig5}), respectively, revealing that a satisfactory tracking performance is achieved in the presence of a $1\%$ error in the estimation. Figure (\ref{fig6}) depicts applied torques generated by the proposed visual-inertial-based control algorithm and confirms that these control signals are realizable and applicable.
	
\section{Conclusion}
The aim of this paper was to address the problem of task-space control of robot manipulators. It is well-known that task-space control laws, which only utilize proprioceptive sensors, are sensitive to inverse kinematic solutions, which are usually associated with model uncertainty and calibration errors. To solve this problem, this paper suggested using exteroceptive vision sensors and introduced a visual-inertial-based control structure. In this control strategy, a hybrid observer was presented for visual-inertial navigation systems, estimating the end-effector pose to provide task-space feedback information. Moreover, a hybrid controller was designed to control the robot manipulator in task-space utilizing the end-effector pose estimated by the proposed hybrid observer. Invoking the Lyapunov stability theorem, it has been proven that the proposed visual-inertial-based control structure is globally asymptotically stable. In the simulation section, the proposed observer-based control strategy has been applied to a 6-DOF long-reach hydraulic manipulator to verify its accuracy and efficiency.
	
{\appendix[Useful Properties of $SE_2(3)$ and $SE(3)$]
Some useful identities and properties of the matrix Lie groups $(SE_2(3), SE(3))$ and the associated Lie algebra are given as follows:
		
\begin{equation*}
\begin{split}
& \text{tr}(\mathcal{V}^T\mathcal{X})=\text{tr}(\mathcal{V}^T\Upsilon(\mathcal{X})), \quad \forall \mathcal{V} \in T_{\mathcal{X}} SE_2(3), \quad (a) \\
& \left\langle \mathcal{V},\mathcal{X} \right\rangle=\left\langle \mathcal{V},\Upsilon(\mathcal{X}) \right\rangle=\left\langle \Upsilon(\mathcal{X}),\mathcal{V} \right\rangle, \quad (b)\\
&\left\langle \mathbb{X},\mathbb{W} \right\rangle= 2\bar{\varphi}(\mathbb{W})^T\bar{\varphi}(\mathbb{X}), \quad (c)\\
& \frac{\partial \text{tr}(A\mathcal{X}B\mathcal{X}^TC)}{\partial \text{tr}(\mathcal{X})}=B\mathcal{X}^TCA+B^T\mathcal{X}^TA^TC^T, \ (d)\\
\end{split}
\end{equation*}
		
\bibliographystyle{IEEEtran}
\bibliography{mybibfile}
		
\begin{IEEEbiography}[{\includegraphics[width=1in,height=1.25in,clip,keepaspectratio]{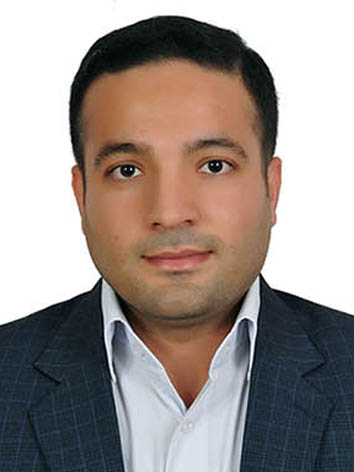}}]{Seyed Hamed Hashemi}
received a B.Sc. in electrical engineering from Babol Noshirvani University of Technology (BNUT), Iran, in 2014; an M.Sc. in electrical engineering-control from Shahrood University of Technology (SUT), Iran, in 2017; and a Ph.D. in electrical engineering-control from Ferdowsi University of Mashhad (FUM), Iran, in 2021. He is currently a postdoctoral research fellow with the Automation Technology and Mechanical Engineering Unit in the Faculty of Engineering and Natural Sciences, Tampere University. His research interests include topological constraints in control systems, the control of hybrid systems, simultaneous localization and mapping, and estimation theory.
\end{IEEEbiography}
		
\begin{IEEEbiography}[{\includegraphics[width=1in,height=1.25in,clip,keepaspectratio]{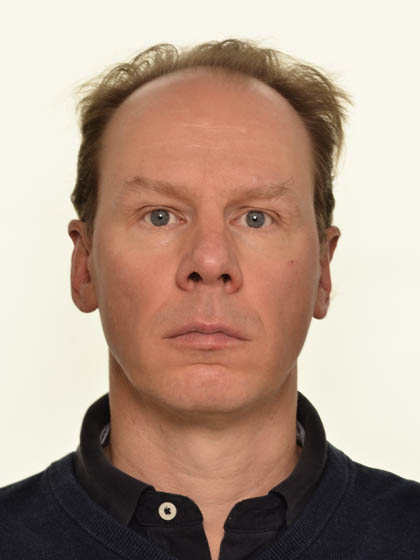}}]{Jouni Mattila}
received his M.Sc. and Ph.D. in automation engineering, both from Tampere University of Technology, Tampere, Finland, in 1995 and 2000, respectively. He is currently a professor of machine automation with the Automation Technology and Mechanical Engineering Unit, Tampere University. His research interests include machine automation, nonlinear model-based control of robotic manipulators, and energy-efficient control of heavy-duty mobile manipulators.
\end{IEEEbiography}
		
\vfill
		
\end{document}